\documentclass[aps,pra,amssymb, amsmath,nobibnotes, nobibnotes,  superscriptaddress, showpacs, showkeys]{revtex4-2}
\usepackage{graphicx}
\usepackage{graphicx,graphics,color,epsfig}
\usepackage{revsymb4-1}
\usepackage{bm}
\usepackage{braket}
\usepackage{mathptmx}
\usepackage[colorlinks, linkcolor=blue, anchorcolor=blue, citecolor=blue]{hyperref}
\usepackage{hyperref}
\usepackage{dsfont}
\usepackage{amsbsy,amsmath}
\usepackage{amssymb}
\usepackage{xcolor}

\DeclareMathAlphabet\mathcal{OMS}{cmsy}{n}{n}
\DeclareMathAlphabet\mathbfcal{OMS}{cmsy}{b}{n}
\newcommand{\bn}[1]{\mbox{\boldmath$#1$}}

\newcommand{\beq}{\begin{equation}}
\newcommand{\eeq}{\end {equation}}
\newcommand{\bea}{\begin{eqnarray}}
\newcommand{\eea}{\end{eqnarray}}
\usepackage{orcidlink}

\begin{document}

\title{Absorption of hybrid fiber modes by Cs atoms in quadrupole transitions}

\author{Smail Bougouffa \orcidlink{0000-0003-1884-4861}}
\thanks{Corresponding author}
\email{sbougouffa@imamu.edu.sa}
\affiliation{Department of Physics, College of Science, Imam Mohammad ibn Saud Islamic University (IMSIU), P.O. Box 90950, Riyadh 11623, Saudi Arabia}

\author{ Mohamed Babiker\orcidlink{0000-0003-0659-5247}}
\thanks{Corresponding author}
\email{m.babiker@york.ac.uk}
\affiliation{Department of Physics, University of York, Heslington, York, YO10 5DD, United Kingdom}

\date{\today}

\begin{abstract}
We evaluate the rate of the absorption of an optical nanofiber mode by a Cs atom in an electric quadrupole transition. With the Cs atom localized near the outer surface of the optical nano-fiber, an interaction occurs between the atomic quadrupole tensor components and the gradients of the vector components of the electric field of a hybrid fiber mode. The absorption rate is evaluated as a function of the radial position of the atom from the fiber axis,  assuming a specific value of the laser power and we use experimentally accessible parameters. We find that the absorption of the hybrid modes by the Cs atom decreases as the atom recedes away from the fiber axis and it formally vanishes at sufficiently large radial distances.  Close to the fiber, however, the absorption rate for the input power chosen can be two orders of magnitude larger than the quadrupole de-excitation rate despite the moderate power used. 
\end{abstract}



\maketitle

\section{Introduction}\label{sec1}

It is well-known that both the rates of emission and absorption of the light by a two-level atom are modified when the atom is localized near the surface of a material object. The presence of the material object modifies the electromagnetic fields with which the two-level atom interacts, leading to significant changes of the rates of emission and absorption \cite{Jacob_2012, Romeira2018a, gu_fainman_2017, yang2017purcell, Ruf2021wd, Rybin2016, Pan2021, ScholzMarggraf, afanasev2017circular, afanasev2016high, Chan2016}. For example, if the atomic transition is dipole-active and the atom is localized between two conductor slabs separated by a sub-wavelength distance, then the emission process can be totally suppressed and so, in the absence of any other influences, the atom remains excited indefinitely. 

The most widely considered two-level system in such problems is assumed to have a dipole-active transition, but atomic systems can have transitions between their energy levels which are dipole-forbidden but quadrupole-allowed, as in the case of Cs, Na, and Rb. Recent studies by both theory and experiment have focused on such atoms \cite{Shibata_2017, Kien2017, LeKien2018, Ray2020a, PhysRevA.62.043818, tojo2005precision, barnett2022}. Quadrupole transitions are, of course,  normally much weaker than dipole transitions, and often their observation requires intense input laser light.  However, it is possible to circumvent the use of intense input light and seek to create situations where the fields are sufficiently intense in well-defined regions of space even though the input power is not high.  Indeed the region close to a material surface can have a high electromagnetic density of states with the energy concentrated in a tiny volume near the surface.   

In this paper, we are concerned with the quadrupole interaction of a two-level atom with the electromagnetic fields outside an optical fiber as the material object. There are a number of reasons why such a physical scenario is novel. Firstly there is the experimental possibility of considering an ultra-thin optical fiber that can be immersed in a cold dilute atomic gas as in the recent experiment by Ray et al \cite{Ray2020a}.  Secondly, the guided modes of the optical nano-fiber introduce new effects due to the twisting of the helical wavefronts associated with the phase function $e^{ip\ell\phi}$ with $p=\pm 1$ and $\ell$ a positive integer. Thirdly the propagation of the electromagnetic fields in an optical fiber is characterized by a chirality in the sense that it distinguishes between axial propagation along the $+z$ and the $-z$ axes.  We seek to determine how these features impact the quadrupole interaction of the nano-fiber modes with the two-level atom which is localized outside its surface. We focus on the absorption process involving the quadrupole transition of a Cs atom which is assumed to be localized in the vicinity of the fiber surface.  The aim is to evaluate the upward transition rate for a specific power of the input laser needed to excite the fiber mode.

Figure 1 schematically presents the $`$optical fiber+atom$'$ system where the two-level atom of transition frequency $\omega_a$ is situated outside the fiber at ${\bf R}=(\rho,0,0)$ where $\rho\geq a$, with $a$ the radius.  This problem is essentially a two-center problem with two sets of coordinate systems separated by the radial vector ${\bn {\rho}}$. We assume that the fiber mode propagating along its axis in the +z-direction has a z component of angular momentum denoted by ${\cal L}^{fiber}_z$, say, relative to the fiber frame of reference. However, relative to the atomic frame the z-component of the angular momentum is  ${\cal L}_z^{atom}$ which is given by
\beq 
{\cal L}_z^{atom}={\cal L}_z^{fiber}-[{\mathbf{\rho}}\times {\bar {\bn {\pi}}}]_z={\cal L}_z^{fiber}-\frac{1}{2}\epsilon_0\rho S_{\varphi}\label{ell}
\eeq
where we have defined ${\bar {\bn {\pi}}}=\frac{1}{c^2}{\bf S}$ as the linear momentum density, ${\bf S}=\frac{1}{2}\Re[{\bf E}^*\times{\bf H}]$ the Poynting vector with ${\bf E}$ and ${\bf H}$, the electric and magnetic fields of the fiber mode.  On applying the results of the analysis to a particular fiber mode and a particular atomic transition we have to determine whether the second term is sufficiently small to write
\beq
{\cal L}_z^{atom}\approx {\cal L}_z^{fiber}\label{ell0}
\eeq
which means that for interactions at the atomic position the optical angular momentum of the mode differs very little from that relative to the fiber frame of reference.  In Appendix B we evaluate the cylindrical components of the Poynting vector of the specific fiber mode and conclude that the azimuthal component of the Poynting vector, $S_{\varphi}$, is small in the region outside the fiber and the near equality in Eq.(\ref{ell0}) is reasonably well justified for the parameters chosen.  The above argument was put forward by Berry \cite{berry1998} in the context of free space optical vortex modes, but here we had to check and confirm its validity for fiber modes.  

The flow of this paper is as follows. In Sec. \ref{sec2} we describe the relevant hybrid modes of the optical fiber and we focus on the Cs quadrupole transitions 
$\ket {L=0,m_l=0}\rightarrow \ket{L=2, m_l'}$ where 
$m_l’=0,\pm 1, \pm 2$. 
In Sec. \ref{sec3} we present the general formalism of the emitter as a two-level system interacting with an external optical field through an electric quadrupole transition. In Sec. \ref{sec4} analytical expressions are derived for the quadrupole Rabi frequency associated with the absorption of the light in a quadrupole transition of a  Caesium atom.  
Section \ref{sec5} is concerned with the absorption process when the atom interacts with the optical fiber field at near-resonance, with the aim of evaluating the absorption rate. The model treats the atom as a two-level system and applies the Fermi Golden rule involving the density of states with appropriate use of the transition selection rules. In presenting the rate of transition formalism we describe the density of the continuum final states as a Lorentzian function representing the upper atomic level as an energy band of width $\hbar \gamma$ where $\gamma^{-1}$ is the free space lifetime of the upper state involved in the quadrupole transition. We adopt experimentally accessible parameters to evaluate the magnitude of the absorption rate.  
In \ref{sec6} we summarise and outline the main conclusions of this paper. 

\section{Hybrid nano-fiber modes} \label{sec2}

\begin{figure}[tbh]
\centering
 \includegraphics[width=0.5\linewidth,height=0.5\linewidth]{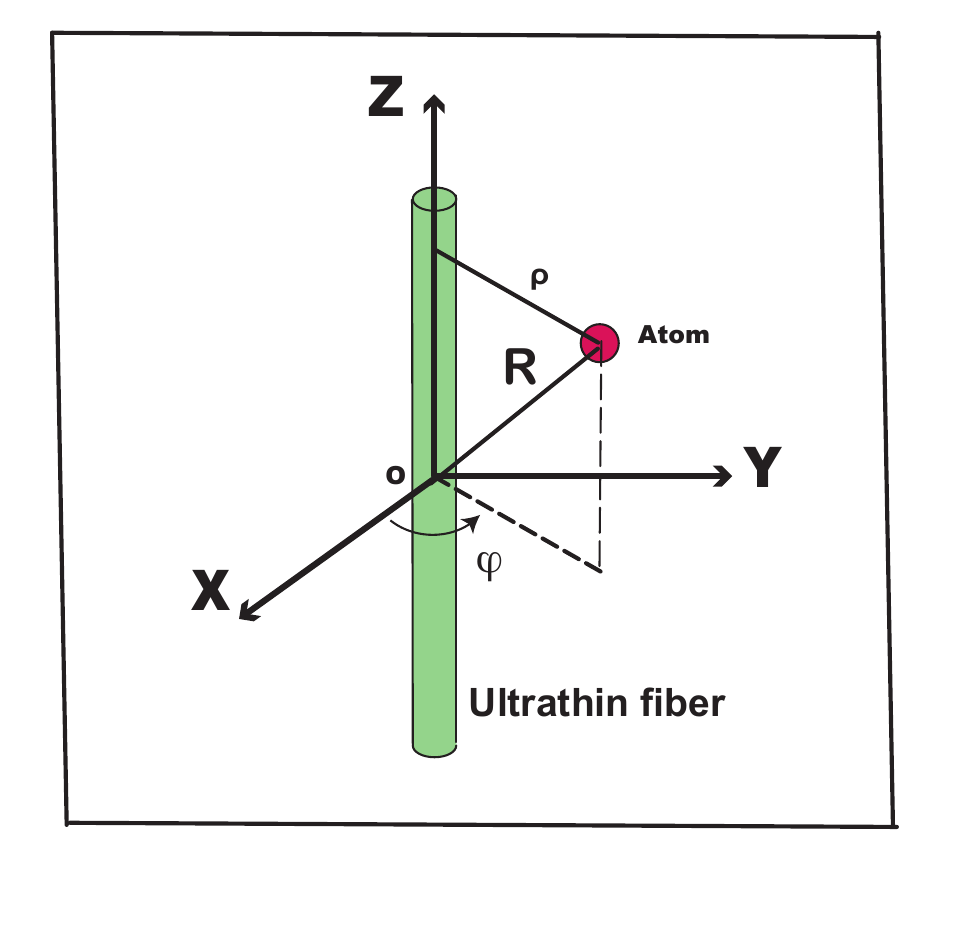}
 \caption{(Color online) An atom, as a two-level system, localized at the position vector ${\bf R}$ in the vicinity of the optical fiber where use is made of both the  Cartesian coordinate system ${\bf R}=\{X, Y, Z\}$ and cylindrical polar coordinates ${\bf R}=(\rho,\varphi,Z)$. The refractive indices in the core $n_1$ and in the cladding regions $n_2$ are both constants.}
\label{Fig1}
\end{figure}

The electromagnetic fields with which the emitter interacts are those in the vicinity of the surface of a nano-fiber in the form of a long circular solid cylinder as shown in Fig. \ref{Fig1} with the fiber surrounded by a homogeneous medium (cladding). The refractive indices in the core region, $n_1$, and in the cladding region, $n_2$, are both assumed to be constants. The theory of fiber modes is described in detail in Refs. \cite{marcuse1982light, snyder2012optical,okamoto2006fundamentals}, so only a brief mention of the modes and more details pertaining to the hybrid modes are presented here. 

 Since the refractive indices in the core and the cladding regions are both constants the fiber can only support guided modes. Thus the types of guided modes that can be excited in the optical fiber are quasi-circularly polarized hybrid modes (HE or EH), transverse electric modes (TE), and transverse magnetic modes (TM). For ease of notation, the fiber guided modes of frequency $\omega$ and axial wavenumber $k$ will now be labeled by the index $\alpha = \{\omega, C, s, p\}$, with the label $C = HE_{\ell m} , EH_{\ell m} , TE_{0m}$ , or $TM_{0m}$ representing the mode kind of integer $\ell = 0,1,2, . . .$ and $m = 1,2, . . .$, with $\ell$  the winding number and $m$  the radial index, respectively.  The index $s = - 1$ or $+1$ denotes the backward or forward propagation along the fiber axis $z$, and $p=\pm 1$ denotes the polarization index. As pointed out earlier, we adopt cylindrical polar coordinates to represent the position vector variable of the center of mass ${\bf R}$ of the atom, so that ${\bf R}=(\rho, \varphi, Z)$.  We write for the electric field of a guided mode 
\begin{equation}
\label{ e1}
\mathbf{E}({\bf R},t)=\mathbfcal{E }(\rho)e^{i\theta^{\alpha}}e^{-i\omega t}+\mathrm{c.c.}
\end{equation}
where $\theta^{\alpha}=s\beta Z+p \ell\varphi$ is the phase function with $\beta$ the propagation constant.  In most standard treatments of the guided modes of an optical fiber the vector amplitude function $\mathbfcal{E}(\rho)$ is normally derived using cylindrical polar coordinates.  It has three components, namely radial, azimuthal, and axial components, and is written as $(\mathcal{E}_\rho, \mathcal{E}_\varphi, \mathcal{E}_Z)$. The quasi-circularly polarized hybrid mode \cite{marcuse1982light, snyder2012optical, okamoto2006fundamentals, Kien2017,kapoor2000mode} is the type of mode of particular importance here.

\subsection{Quasi-circularly polarized hybrid modes}

For integer $\ell > 0$ we have the quasi-circularly polarized hybrid modes $C= $\{HE$_{\ell m}$ or EH$_{\ell m}\}$ and are defined both inside the fiber $(\rho<a)$ and outside it $(\rho>a)$. The components of the electric field inside the fiber $\rho < a$ are as follows

\begin{eqnarray}\label{e2}
\mathcal{E}_{\rho}&=& i\mathcal{N}\frac{\beta}{2\mu}[(1-\xi)J_{\ell-1}(\mu \rho)-(1+\xi)J_{\ell+1}(\mu \rho],\nonumber\\
\mathcal{E}_{\varphi}&=& -\mathcal{N}\frac{\beta}{2\mu}[(1-\xi)J_{\ell-1}(\mu \rho)+(1+\xi)J_{\ell+1}(\mu \rho)],\nonumber\\
\mathcal{E}_{Z}&=& \mathcal{N}J_{\ell}(\mu \rho), 
\end{eqnarray}
and, for $\rho>a$,
\begin{eqnarray}\label{e3}
\mathcal{E}_{\rho}&=& i\mathcal{N}\frac{\beta}{2\nu}\frac{J_{\ell}(\mu a)}{K_{\ell}(\nu a)}[(1-\xi)K_{\ell-1}(\nu \rho)+(1+\xi)K_{\ell+1}(\nu \rho)],\nonumber\\
\mathcal{E}_{\varphi}&=&-\mathcal{N}\frac{\beta}{2\nu}\frac{J_{\ell}(\mu a)}{K_{\ell}(\nu a)}[(1-\xi)K_{\ell-1}(\nu \rho)-(1+\xi)K_{\ell+1}(\nu \rho)],\nonumber\\
\mathcal{E}_{Z}& = & \mathcal{N}\frac{J_{\ell}(\mu a)}{K_{\ell}(\nu a)}K_{\ell}(\nu \rho).
\end{eqnarray}
where $\xi$ is a system parameter, defined as 
\begin{equation}\label{e4}
\xi=l\left(\frac{1}{\mu^2a^2}+\frac{1}{\nu^2a^2}\right)\left[\frac{J_{\ell}'(\mu a)}{\mu aJ_{\ell}(\mu a)}
+\frac{K_{\ell}'(\nu a)}{\nu aK_{\ell}(\nu a)} \right]^{-1},
\end{equation}
where the prime stands for the total derivative. In the above field components, $\mu=(n_1^2k^2-\beta^2)^{1/2}$ is the wave number associated with the radial variation of the field inside the fiber, and $\nu=(\beta^2-n_2^2k^2)^{1/2}$ is associated with the spatial decay of the field amplitude radially outside the fiber. 
The functions $J_n$ and $K_n$, with $n$ integer, are Bessel functions of the first kind and modified Bessel functions of the second kind, respectively. Finally, $\mathcal{N}$ is the overall constant which is determined in terms of the power ${\cal P}$ of the field.  The evaluation of $|\mathcal{N}|$ is described in Appendix A in which we find $|{\cal N}|$ is given by
\begin{equation}
|\mathcal{N}|^2=\frac{{\mathcal{P}}}{{\cal I}_{H1}+{\cal I}_{H2}}
\end{equation}
where for the ${\cal I}_{H1}$ and ${\cal I}_{H2}$ are the integrals appearing in the Appendix as in Eqs.(\ref{e13p}) and (\ref{e14p}) and are to be evaluated numerically.

The set of equations (\ref{e2}) and ( \ref{e3}) which display the electric field components of the fiber show that the fiber field acquires not only two individual transverse (radial and azimuthal) components but also a longitudinal (axial) component.  The existence of the common phase factor $e^{i \ell \varphi}$ in Eq.\ref{ e1} means that there is an azimuthal-phase dependence. Such a phase dependence is characteristic of the field in fiber modes and is directly responsible for the orbital angular momentum content of the mode.

\begin{figure}[h]
\centering
\includegraphics[width=0.33\linewidth,height=0.3\linewidth]{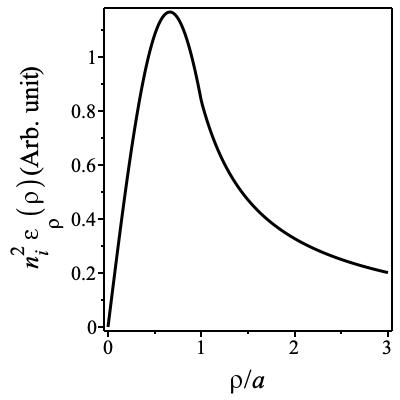}~\includegraphics[width=0.33\linewidth,height=0.3\linewidth]{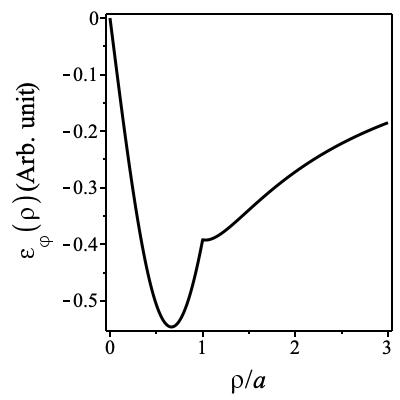}~\includegraphics[width=0.33\linewidth,height=0.3\linewidth]{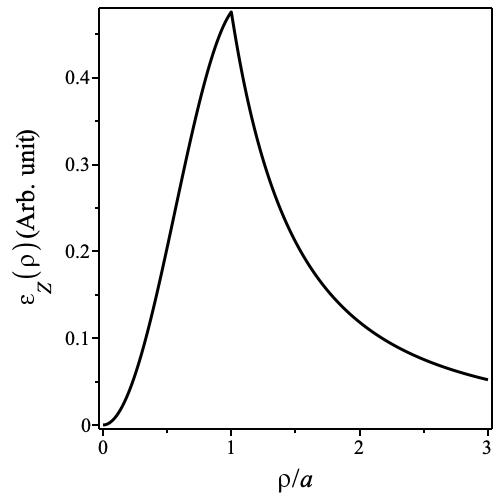} \\
\includegraphics[width=0.33\linewidth,height=0.3\linewidth]{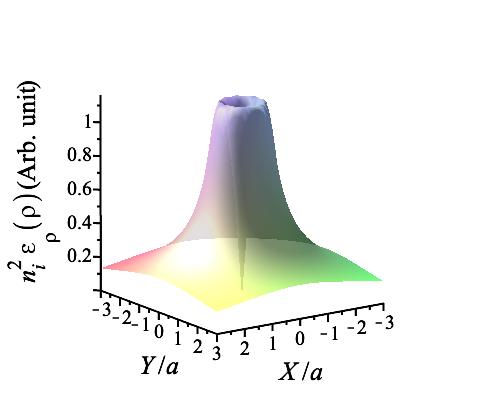}~\includegraphics[width=0.33\linewidth,height=0.3\linewidth]{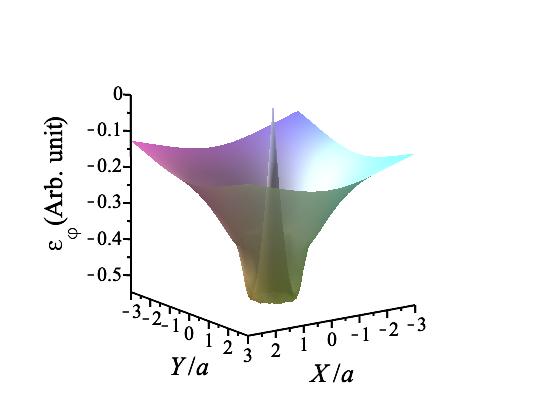}~\includegraphics[width=0.33\linewidth,height=0.3\linewidth]{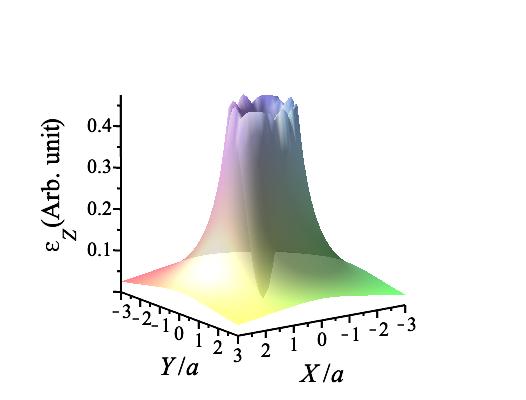} 
\caption{(Color online) Upper panel: variations of the field components ${\cal E}_\rho,{\cal E}_\varphi,{\cal E}_z$ with the radial coordinate $\rho$ for the $EH $ optical fiber mode for which $(\ell=2, m=1)$. Lower panel: variations in the XY plane indicating cylindrical symmetry. The refractive indices of the fiber core material and the vacuum cladding are $n_1=1.4615$ and $n_2=1$, respectively. The fiber radius is $a=290$ nm and the wavelength of the light is set to $\lambda=685 nm$.}\label{Fig2-1} 
\end{figure}

Figure (\ref{Fig2-1}) displays radial variations of the field components within the core of the fiber ($\rho<a$) and outside it ($\rho>a$) for optical fiber mode $EH$ for which $(\ell=2, m=1)$. The refractive indices of the fiber and the vacuum cladding are $n_1=1.4615$ and $n_2=1$. 

\subsection{Dispersion relation}

For a fiber mode of frequency $\omega$, wavelength  $\lambda=2\pi c/\omega$ and wave number $k=\omega/c$, the propagation constant $\beta$  satisfies the transcendental equation \cite{marcuse1982light}
\begin{eqnarray}\label{e9}
\lefteqn{\bigg[\frac{J_{\ell}'(\mu a)}{\mu aJ_{\ell}(\mu a)}
+\frac{K_{\ell}'(\nu a)}{\nu aK_{\ell}(\nu a)}
\bigg]\bigg[\frac{n_{1}^2J_{\ell}'(\mu a)}{\mu aJ_{\ell}(\mu a)}
+\frac{n_{2}^2K_{\ell}'(\nu a)}{\nu aK_{\ell}(\nu a)}
\bigg]}\nonumber\\&&\mbox{}\qquad\qquad\qquad\qquad
=\Big(\frac{\ell\beta}{k}\Big)^2\left(\frac{1}{\mu^2a^2}+\frac{1}{\nu^2a^2}\right)^2,
\end{eqnarray}

The case $\ell=0$ is concerned with the TE and TM modes, but we are interested here in the case  $\ell \neq 0$, so  Eq.\ref{e9} involves a mixture of HE and EH modes  \cite{marcuse1982light, snyder2012optical,okamoto2006fundamentals}. For HE modes we have
\begin{equation}\label{e9b}
f_{HE}(\beta)=\frac{J_{\ell-1}(\mu a)}{\mu aJ_{\ell}(\mu a)}-\frac{\ell}{\mu^2a^2}+\frac{1}{2}\Big(1+\frac{n_{2}^2}{n_{1}^2}\Big)\frac{K'_{\ell}(\nu a)}{\nu aK_{\ell}(\nu a)}+\mathcal{A}=0
\end{equation}
The dispersion relation for the  EH modes differs by the negative sign multiplying ${\cal A}$ and so on evaluation it gives rise to different values of $\beta$
\begin{equation}\label{e9c}
f_{EH}(\beta)=\frac{J_{\ell-1}(\mu a)}{\mu aJ_{\ell}(\mu a)}-\frac{l}{\mu^2a^2}+\frac{1}{2}\Big(1+\frac{n_{2}^2}{n_{1}^2}\Big)\frac{K'_{\ell}(\nu a)}{\nu aK_{\ell}(\nu a)}-\mathcal{A}=0,
\end{equation}
where $\mathcal{A}$ is given by
\begin{equation}\label{e9d}
\mathcal{A}=\bigg[ \bigg(\frac{\ell\beta}{n_{1}k}\bigg)^2\bigg(\frac{1}{\nu ^2a^2}+\frac{1}{h^2a^2}\bigg)^2+\frac{1}{4}\bigg(1-\frac{n_{2}^2}{n_{1}^2}\bigg)^2\bigg(\frac{K'_{\ell}(\nu a)}{\nu aK_{\ell}(qa)}\bigg)^2\bigg]^{1/2}.
\end{equation}

The HE and EH modes are labeled by HE$_{\ell m}$ and EH$_{\ell m}$, respectively, such that $\ell=1,2,\dots$ and $m=1,2,\dots$ are the azimuthal and radial mode orders, respectively. 
In this case, the radial mode order $m$ indicates that the HE$_{\ell m}$ or EH$_{\ell m}$ mode is the solution to the corresponding equations \ref{e9b} or \ref{e9c}.

\begin{figure}[ht]
\centering
\includegraphics[width=0.3\linewidth,height=0.38\linewidth]{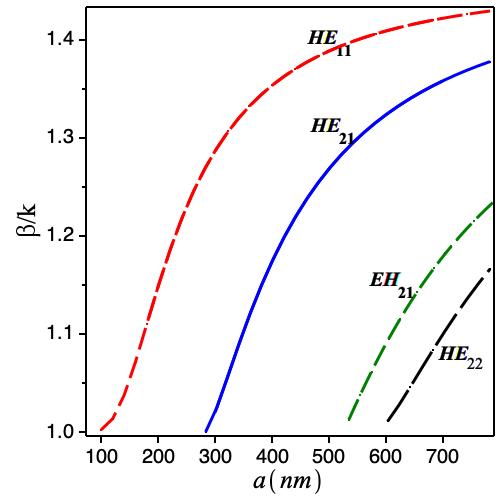}~
\includegraphics[width=0.3\linewidth,height=0.38\linewidth]{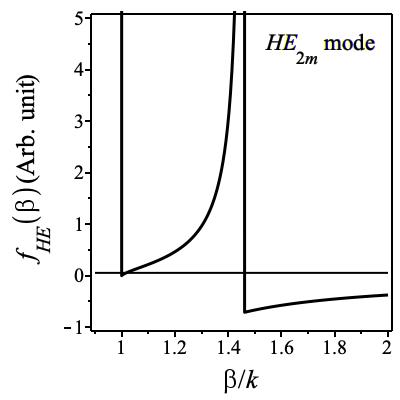}~\includegraphics[width=0.3\linewidth,height=0.38\linewidth]{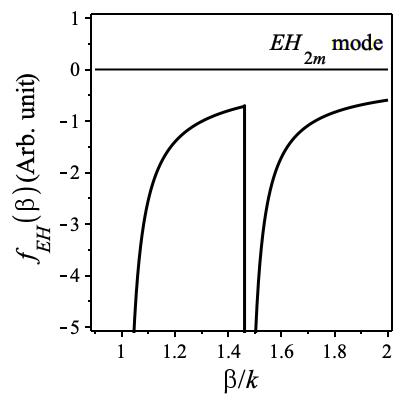} 
\caption{(Color online) Left panel: variations of the scaled propagation constant $\beta/k$ with the fiber radius $a$. The wavelength of the light is set to $\lambda=685 nm$.
Middle and right panels: variations of the dispersion functions  $f_{HE}$ and ${f_{EH}}$ with $\beta/k$ for $\ell=2\;m=1$.  The roots of the dispersion functions determine the values of $\beta$ in each case.  Once again, the refractive indices of the fiber and the vacuum cladding are assumed to be  $n_1=1.4615$ and $n_2=1$, and the fiber radius is $a=290$ nm. }\label{Fig2-2} 
\end{figure}

 Figure (\ref{Fig2-2} -left panel) displays the variations of the propagation constant of the fiber as a function of the fiber radius $a$ at optical wavelength $\lambda= 685$ nm for the fundamental $HE_{11}$ mode as well as some higher-order modes. However, for the optical fiber under consideration here the higher-order mode, $HE_{21}$ is realizable at $a \simeq 283$ nm.  It is well-known that as the fiber radius $a$ increases the number of modes it can support increases. Our numerical results agree with those in the literature \cite{Sague, frawley2012higher, kumar2015interaction, Kien2017}. 

Figures (\ref{Fig2-2}-middle and right panels), display the dispersion functions \ref{e9b} and \ref{e9c} against $\beta/k$ for the $HE, EH$ with $(\ell=2)$  where the fiber radius $a=290$ nm. The refractive indices of the fiber and the vacuum cladding are $n_1=1.4615$ and $n_2=1$, respectively.  

\section{Quadrupole Interaction}\label{sec3}
\subsection{Derivation of the Rabi frequency}

Having described both the mode functions and the dispersion relations of the hybrid fiber modes with which the atomic two-level system interacts, we now turn to describe the two-level system itself as consisting of a two-level atom engaging with the fiber through its electric quadrupole moment. The ground and excited states of the two-level-atom involved in the atomic quadrupole transition are $\{\ket{g}, \ket{e}\}$ with corresponding frequencies $\omega_g$ and $\omega_e$, respectively, corresponding to the resonance frequency $\omega_a=(\omega_e- \omega_g)$. The interaction Hamiltonian is written in Cartesian coordinate systems as a multipolar series with fields evaluated at the center of mass coordinate $\mathbf{R}=(X, Y, Z)$ and can be written as  \cite{Bougouffa, Bougouffa20, babiker2018atoms, lembessis2013enhanced, Al-Awfi2019, Bougouffa_2021}
\begin{equation}\label{e10}
    \hat{H}_{int}=\hat{H}_{D}+\hat{H}_{Q}+...,
\end{equation}
where the first term  $\hat{H}_{D}=- \hat{\bm{d}}.\mathbf{ \hat{E}}(\mathbf{R})$  stands for the electric dipole interaction between the neutral atom and the electric field,  
$\bm{ \hat{d}}$ is the electric dipole moment vector, where $\mathbf{r}=(x,y,z)$ is the internal electronic-type position vector with components $(x,y,z)$ written as $x_i,\;\;i=1,2,3$. $\bm{ \hat{E}}(\mathbf{R})$ is the electric field vector evaluated at the center of mass coordinate ${\bf R}$. The atomic transition process in question is taken here to be dipole-forbidden but quadrupole-allowed, so it is the second (quadrupole) interaction term in Eq.(\ref{e10}) that dominates in this case.  We have

\begin{equation}\label{e11}
 \hat{H}_{Q}=-\frac{1}{2}\sum_{ij} \hat{Q}_{ij} \nabla_i \hat{E_j}.
\end{equation}
This is essentially the interaction involving the Cartesian components of the quadrupole moment tensor ${\hat{Q}}_{ij}=ex_ix_j$ and the gradients of the electric field vector components, evaluated at the center-of-mass coordinate $\mathbf{R}$.  
Here $\nabla_i$ are components of the ${\bn {\nabla}}$ vector operator which act only on the spatial coordinates of the transverse electric field vector $\mathbf{ E}$ as a function of the center of mass vector $\mathbf{ R}= (X, Y, Z)$.

For the two-level atom, the quadrupole tensor operators  ${\hat{Q}}_{ij}$  can be written in terms of ladder operators as $ \hat{Q}_{ij}=Q_{ij}( \hat{b} + \hat{b}^{\dag})$, where $Q_{ij}=\bra{i}\hat{Q}_{ij}\ket{j}$ are the quadrupole matrix elements between the two atomic levels, and $ \hat{b} ( \hat{b}^{\dag})$ are the atomic level lowering (raising) operators. The electric field Eq. (\ref{ e1}) can now be written in the Cartesian form
\begin{equation}\label{e12}
    \mathbf{ E}(\mathbf{R})=\sum_{i}\mathbf{ \hat{e}_{i}}E_i,
\end{equation}
where $\mathbf{ \hat{e}_{i}} (i=x,y,z)$ are the Cartesian unit vectors and $E_i$ are the Cartesian components of the optical electric field that can be written as
\begin{equation}\label{e13}
E_i= u^{\{\alpha\}}_i(\mathbf{R})e^{i \theta^{\{\alpha\}} (\mathbf{R})}e^{-i\omega t}+ c.c. , 
\end{equation}
where c.c. stands for complex conjugate;  $u^{\alpha}_i(\mathbf{R})$ and $\theta^{\alpha} (\mathbf{R})$ are the amplitude and the phase functions of the Cartesian $i^{th}$ optical electric field component. Recall that the superscript $\alpha $ denotes a group of indices that specify the optical mode in terms of its frequency $\omega$, the mode kind $C$, azimuthal and radial numbers $\ell$ and $m$, and the polarization index $p$.  The quadrupole interaction Hamiltonian can now be written as follows

\begin{equation}\label{e14}
  \hat{H}_{Q}=-\frac{1}{2}\sum_{i,j} \hat{Q}_{ij}\frac{\partial {E_j}}{\partial R_i}.
\end{equation}
and this interaction Hamiltonian can also be written in terms of the Rabi frequency as follows
\begin{equation}\label{e15}
    \hat{H}_{Q}=-\hbar\Omega_{Q}^{\{\alpha\}}(\mathbf{R})\hat{a}e^{i\theta^{\{\alpha\}}(\mathbf{R})}e^{-i\omega t}+H.c.
\end{equation}
where ${\hat a}$ and ${\hat a}^{\dagger}$ are the fiber mode destruction and creation operators and $\Omega_{Q}^{\{\alpha\}}(\mathbf{R})$ is the quadrupole Rabi frequency

\begin{equation}\label{e16}
   \Omega_{Q}^{\{\alpha\}}(\mathbf{R})=\frac{1}{2\hbar}\sum_{i,j} Q_{ij} u^{\{\alpha\}}_j \Big( \frac{1}{ u^{\{\alpha\}}_j}\frac{\partial u^{\{\alpha\}}_j}{\partial R_i}+i\frac{\partial \theta^{\{\alpha\}}}{\partial R_i} \Big)
\end{equation}
This is the general form of the quadrupole Rabi frequency which applies to any of the allowed modes.

Recall that the optical fiber is a long dielectric cylinder of radius $a$ and refractive index $n_1$ immersed in a background medium of refractive index $n_2$, where $n_2<n_1$. The Quadrupole interaction is expressed in terms of the Cartesian coordinates $\{x, y, z\}$ relative to the atomic centre of mass, with the centre of mass coordinate written ${\bf R}=(X, Y, Z)$.  The amplitude functions of the electric field components of the fiber modes ${\mathcal {E}}_j$ are, however, given in cylindrical coordinates ${\bf R}=(\rho,\varphi, Z)$, so to proceed we need to make a transformation to express the optical electric field components of the fiber modes in terms of $ u^{\{\alpha\}}_j({\bf R})$.  We have
\begin{eqnarray}\label{e17}
u^{\{\alpha\}}_{x}&=& \cos(\varphi)\mathcal{E}_\rho-\sin(\varphi)\mathcal{E}_{\varphi},\nonumber\\ 
u^{\{\alpha\}}_{y}&=&\sin(\varphi)\mathcal{E}_\rho+\cos(\varphi)\mathcal{E}_{\varphi},\nonumber\\  
u^{\{\alpha\}}_{z}&=& \mathcal{E}_Z,
\end{eqnarray}
where $\theta^{\{\alpha\}}=(s\beta Z+p \ell\varphi)$.  Also $\rho=\sqrt{X^2+Y^2}$ and $\varphi=\tan^{-1}(Y/X)$.  
The quadrupole Rabi frequency Eq. (\ref{e16}) can now be written as the sum of three terms
\begin{equation}\label{e18}
   \Omega_{Q}^{\{\alpha\}}(\mathbf{R})=\frac{1}{2\hbar}\sum_{j=1}^{3}\Omega_j,
\end{equation}
where
\begin{eqnarray}\label{e19}
 \Omega_{1}&=&\big( \frac{\partial u^{\{\alpha\}}_{x} }{\partial X} -i \xi_1 u^{\{\alpha\}}_{x} \big)Q_{xx}+\big( \frac{\partial u^{\{\alpha\}}_{y} }{\partial X} -i \xi_1 u^{\{\alpha\}}_{y} \big)Q_{xy}+\big( \frac{\partial u^{\{\alpha\}}_{z} }{\partial X} -i \xi_1 u^{\{\alpha\}}_{z} \big)Q_{xz} ,\nonumber\\ 
 \Omega_{2}&=&\big( \frac{\partial u^{\{\alpha\}}_{x} }{\partial Y} +i \xi_2 u^{\{\alpha\}}_{x} \big)Q_{yx}+\big( \frac{\partial u^{\{\alpha\}}_{y} }{\partial Y} +i \xi_2 u^{\{\alpha\}}_{y} \big)Q_{yy}+\big( \frac{\partial u^{\{\alpha\}}_{z} }{\partial Y} +i \xi_2 u^{\{\alpha\}}_{z} \big)Q_{yz} ,\nonumber\\  
 \Omega_{3}&=& is\beta\big( u^{\{\alpha\}}_{x}Q_{zx}+u^{\{\alpha\}}_{y}Q_{zy}+u^{\{\alpha\}}_{z}Q_{zz}\big),
\end{eqnarray}

where $\xi_1=p\ell Y/(X^2+Y^2)$ and  $\xi_2=p\ell X/(X^2+Y^2)$. For an absorption transition, the quadrupole Rabi frequency depends on the type of mode excited in the optical fiber involved in the quadrupole transition and the angular momentum quantum numbers of the two energy levels, and these are governed by the transition selection rules.
\subsection{Applying the Selection Rules}

We focus specifically on the quadrupole transition $\ket {L=0,m_l=0}\rightarrow \ket{L=2, m_l'}$ in which a fiber mode is absorbed, where $(m_l'=0, \pm 1, \pm 2)$ and we are adopting the notation $\ket {L, m_{l}}$ for the atomic state, labeled by the angular momentum quantum numbers \cite{Bougouffa, Bougouffa20, babiker2018atoms, lembessis2013enhanced}. 

The selection rules are $\Delta L=0,\pm 2;\;\Delta m=0,\pm 1,\pm 2$.  We consider the following three situations
\begin{itemize}
  \item For the case $m_l'=0$, the quadrupole moment tensor can be evaluated \cite{Bougouffa, Bougouffa20} and we find that all the off-diagonal quadrupole tensor components are equal to zero  $(Q_{xy} = Q_{xz}=Q_{yz}=0)$, while the diagonal matrix elements are non-zero.  We have   $(Q_{xx}=Q_{yy}=Q_0$ and $Q_{zz}=-2Q_0)$. Thus, the Rabi frequency equation (\ref{e18}) takes the following simpler form,
\begin{equation}\label{e20a}
\Omega_{Q}^{\{\alpha\}}(\mathbf{R})=\frac{Q_0}{2\hbar}\Big [ \big( \frac{\partial u^{\{\alpha\}}_{x} }{\partial X} -i \xi_1 u^{\{\alpha\}}_{x} \big)+ \big( \frac{\partial u^{\{\alpha\}}_{y} }{\partial Y} +i \xi_2 u^{\{\alpha\}}_{y} \big) - 2 is\beta u^{\{\alpha\}}_{z}\Big].
\end{equation}
Equation (\ref{e20a}) can be written entirely in terms of the components of $\mathbfcal {E}$ with cylindrical coordinates.  We find
\begin{equation}\label{e20b}
   \Omega_{Q}^{\{\alpha\}}(\mathbf{R})=\frac{Q_0}{\hbar}\Big [ \frac{1}{\rho} \frac{\partial (\rho\mathcal{E}_\rho)}{\partial \rho} +\frac{ip\ell }{\rho}\mathcal{E}_{\varphi} - 2 is\beta \mathcal{E}_{Z}\Big].
\end{equation}

  \item For the case $m_l'=\pm 1$, the diagonal matrix elements are zero  $(Q_{xx}=Q_{yy}=Q_{zz}=0)$ and the off-diagonal matrix elements are $Q_{xy}=0$ and $Q_{xz}=iQ_1, Q_{yz}=\mp Q_1$. Consequently, the Rabi frequency equation (\ref{e18}) yields,
\begin{eqnarray}\label{e21}
   \Omega_{Q}^{\{\alpha\}}(\mathbf{R})=&\frac{Q_1}{2\hbar}\Big [i\big( \frac{\partial u^{\{\alpha\}}_{z} }{\partial X} -i \xi_1 u^{\{\alpha\}}_{z} \big)\mp \big( \frac{\partial u^{\{\alpha\}}_{z} }{\partial Y} +i \xi_2 u^{\{\alpha\}}_{z} \big)\nonumber  \\
   &+is\beta\big(i u^{\{\alpha\}}_{x}\mp u^{\{\alpha\}}_{y}\big)\Big],
\end{eqnarray}  

or in terms of the cylindrical components, we have for the case $m_l'=\pm 1$
\begin{equation}\label{e21b}
   \Omega_{Q}^{\{\alpha\}}(\mathbf{R})=\frac{iQ_1 e^{\pm i \varphi}}{\hbar}\Big [ \frac{\partial \mathcal{E}_Z}{\partial \rho} \mp \frac{ip\ell }{\rho}\mathcal{E}_{Z}  +is\beta\big( \mathcal{E}_\rho\pm i \mathcal{E}_{\varphi}\big)\Big].
\end{equation}  
   
  \item  For the case $m_l'=\pm 2$, the zero matrix elements are  $(Q_{zz}=Q_{yz}=Q_{xz}=0)$ and the non zero matrix elements are $Q_{xx}=-Q_{yy}=Q_1, Q_{xy}=\pm iQ_1$. Accordingly, the Rabi frequency equation (\ref{e18}) has the following form,
\begin{eqnarray}\label{e22}
 \Omega_{Q}^{\{\alpha\}}(\mathbf{R})=&\frac{Q_1}{\hbar}\Big [ \big( \frac{\partial u^{\{\alpha\}}_{x} }{\partial X} -i \xi_1 u^{\{\alpha\}}_{x} \big)-\big( \frac{\partial u^{\{\alpha\}}_{y} }{\partial Y} +i \xi_2 u^{\{\alpha\}}_{y} \big)\nonumber \\
  & \pm i \Big(\big( \frac{\partial u^{\{\alpha\}}_{y} }{\partial X} -i \xi_1 u^{\{\alpha\}}_{y} \big)+\big( \frac{\partial u^{\{\alpha\}}_{x} }{\partial Y} +i \xi_2 u^{\{\alpha\}}_{x}\big) \Big)\Big],
\end{eqnarray}  

and in terms of the cylindrical components, we have for the case $m_l'=\pm 2$
\begin{equation}\label{e22b}
   \Omega_{Q}^{\{\alpha\}}(\mathbf{R})=\frac{Q_1e^{\pm 2i\varphi}}{\hbar}\Big [ \big( \frac{\partial }{\partial \rho}(\mathcal{E}_\rho\pm i\mathcal{E}_{\varphi}) -\frac{1}{\rho}(1\pm p\ell)(\mathcal{E}_\rho\pm i\mathcal{E}_{\varphi})\Big].
\end{equation} 

\end{itemize}
We thus have in hand closed expressions for the Rabi frequencies in the hybrid modes of the optical fiber for the quadrupole transitions  $\ket {L=0,m_l=0}\rightarrow \ket{L=2, m_l'}$ which satisfy the transition selection rules.

\section{Caesium atoms Outside Optical fiber}\label{sec4}

 Once again we emphasize that we are investigating the specific case of the Cs atom, which has also been the subject of an investigation involving its quadrupole transition $(6^2S_{1/2}\rightarrow5^2D_{5/2})$ in various contexts \cite{tojo2005precision, chan2019coupling, lembessis2013enhanced, Shibata_2017, sakai2018nanofocusing}. We have the following as parameters for Cs:  $\lambda=685$ nm, $Q_{xx}=10e a_0^2$ (with $a_0$ the Bohr radius), and the de-excitation rate involved in the quadrupole transition $\gamma=7.8\times 10^5 (s^{-1})$.  The fiber radius is $a=290$ nm and the refractive indices of the fiber core and the vacuum cladding are $n_1=1.4615$ and $n_2=1$, respectively. The overall constant ${\cal N}$ depends on the power ${\cal P}$ of the excited fiber mode and here we set the value of  ${\cal P}$ as  ${\cal P}=2.5 (\mu W)$ \cite{Ray2020a}. 

In Appendix B we evaluate the components of the Poynting vector and exhibit their variations with the atom position $\rho\geq a$ outside fiber. It is clear that the azimuthal component is much smaller than the axial component and we conclude that Eq.(\ref{ell}) is reasonably well satisfied.  We may now consider the processes of absorption of the hybrid optical fiber modes by the atom, having obtained expressions for their Rabi frequencies in preparation for evaluating the absorption rate.  

\subsection{Absorption of an HE mode}

An $HE$ mode is appropriate for the quadrupole transition $\ket {L=0,m_l=0}\rightarrow \ket{L=2, m_l'}$ for the mode $C=$ \{HE$_{\ell m}\}$.  We need not substitute $\ell=2$ at this stage, then using  Eq. (\ref{e3}) and assuming $p=1$, we need to evaluate the quadrupole Rabi frequencies for the different situations.   

For the case $m_l'=0$, Eq.(\ref{e20b}) becomes

\begin{equation}\label{e25a}
   \Omega_{Q}^{\{\alpha\}}(\mathbf{R})=-i\frac{Q_0}{2\hbar}\beta(2s+1)\mathcal{N}\frac{J_{\ell}(\mu a)}{K_{\ell}(\nu a)}K_{\ell}(\nu \rho)
\end{equation}

For the case $m_l'=\pm 1$, Eq.(\ref{e21b}) reads

\begin{equation}\label{e25b}
\Omega_{Q}^{\{\alpha\}}(\mathbf{R})=\pm i \frac{Q_1}{2\hbar}\mathcal{N} e^{\pm i \varphi}\frac{J_{\ell}(\mu a)}{K_{\ell}(\nu a)}\left((i\pm 1)\frac{\ell}{\rho} K_{\ell}(\nu \rho)+\frac{s\beta^2(\xi\pm1)\pm \nu^2}{\nu}K_{\ell+1}(\nu \rho)\right ).
\end{equation}
and we note the explicit appearance of the index $s=\pm 1$, the direction of polarization in the above expressions.

For the case $m_l'=\pm 2$ we obtain from Eq.(\ref{e22b})

\begin{equation}\label{e25c}
\Omega_{Q}^{\{\alpha\}}(\mathbf{R})=
    \mp i \frac{Q_1}{2\hbar}\beta(\xi\pm 1)\mathcal{N} e^{\pm i 2\varphi}\frac{J_{\ell}(\mu a)}{K_{\ell}(\nu a)}\left(K_{\ell}(\nu \rho)+\frac{2(1 \pm \ell)}{\nu \rho}K_{\ell\pm 1}(\nu \rho) \right),
\end{equation}
which does not depend on $s=\pm 1$.

We have to include the Clebsch-Gordan coefficients (CGC) in the formalism since the atomic levels involve both the orbital angular momentum and the spin of the electron, both of which are required in fine structure. The following Clebsch-Gordan coefficients

\begin{eqnarray}\label{17ab}
   CGC &=&\sqrt{\frac{5}{5}}, \quad \sqrt{\frac{4}{5}}, \quad \sqrt{\frac{3}{5}}, \quad \sqrt{\frac{2}{5}}, \quad \sqrt{\frac{1}{5}},
\end{eqnarray}
correspond to the transitions for which $ \Delta m =-2, -1,  0, +1,  +2 .$, respectively \cite{Bougouffa_2021,afanasev2018experimental}.

\begin{figure}[htbp]
\centering
\includegraphics[width=0.19\linewidth,height=0.3\linewidth]{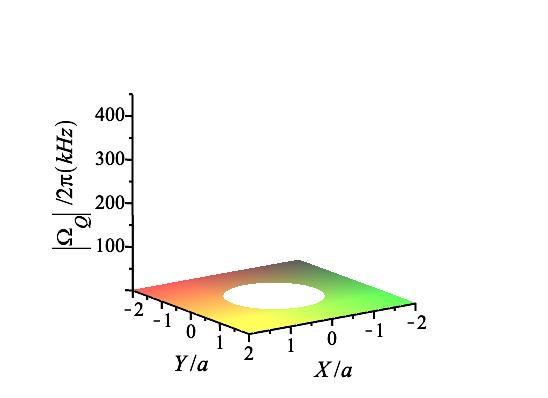}~\includegraphics[width=0.19\linewidth,height=0.3\linewidth]{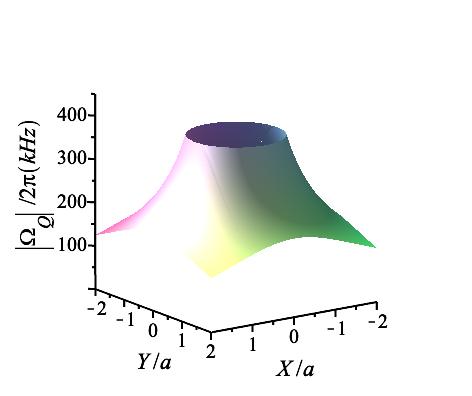}
~\includegraphics[width=0.19\linewidth,height=0.3\linewidth]{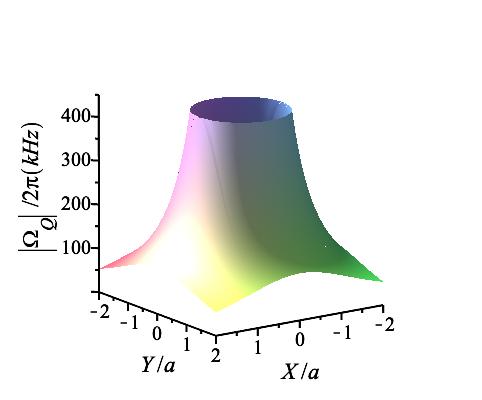}~\includegraphics[width=0.19\linewidth,height=0.3\linewidth]{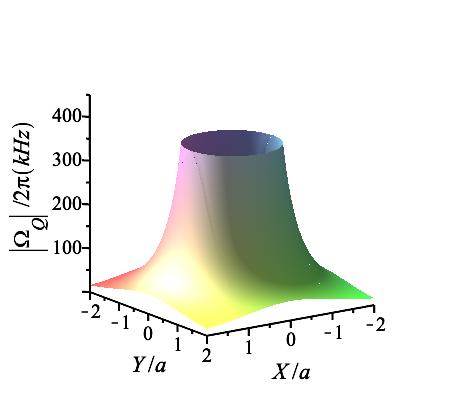}~\includegraphics[width=0.19\linewidth,height=0.3\linewidth]{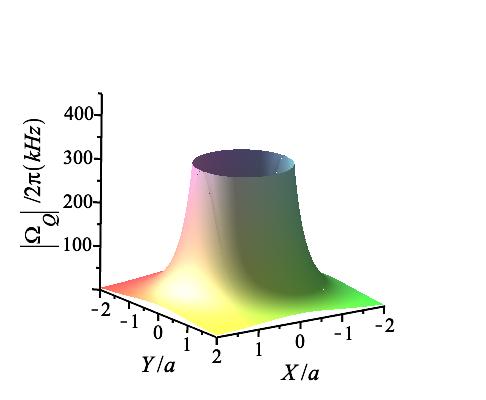}
\caption{(Color online) The spatial distribution of the quadrupole Rabi frequency $|\Omega_Q|/2\pi (kHz)$ for the Cs atom with the quadrupole transition $\ket {L=0,m_l=0}\rightarrow \ket{L=2, m_l'}$ in the mode $C=$ \{HE$_{2 1 }\}$.  For $ m_l'=-2$, $ m_l'=-1$, $ m_l'=0$, $ m_l'=+1$, and $ m_l'=+2$ from the left to the right, respectively. In all cases $s=+1, p=+1$. The power is ${\cal P} = 2.5$($\mu$W ); for other parameters see the text.}\label{Fig4} 
\end{figure}

Figure \ref{Fig4} displays the spatial distribution of the quadrupole Rabi frequency  $|\Omega_Q/2\pi (kHz)|$ for the Cs atom with the quadrupole transition $\ket {L=0,m_l=0}\rightarrow \ket{L=2, m_l'}$ in the mode $C=$ \{HE$_{21}\}$ where $\ell=2$ for different values of $m_l'=0, \pm1, \pm 2$. The results show that the quadrupole Rabi frequency has cylindrical symmetry and exhibits the usual characteristic dependence of decaying amplitude with the radial distance outside the fiber $(\rho>a)$. It is clear from these results that the largest magnitude of the quadrupole Rabi frequency corresponds to the case of the quadrupole transition $\ket {L=0,m_l=0}\rightarrow \ket{L=2, m_l'=0}$. 

\begin{figure}[htbp]
\centering
\includegraphics[width=0.3\linewidth,height=0.3\linewidth]{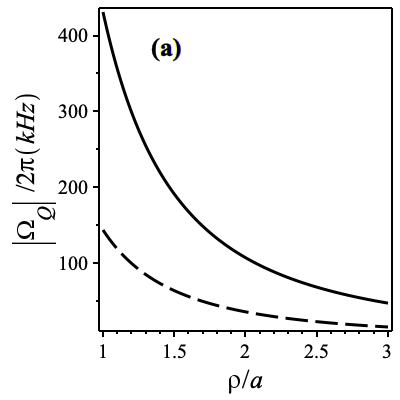}~\includegraphics[width=0.3\linewidth,height=0.3\linewidth]{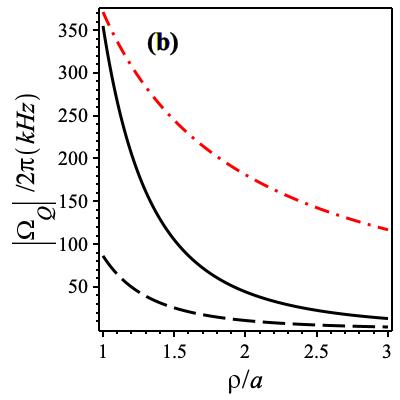} 
~\includegraphics[width=0.3\linewidth,height=0.3\linewidth]{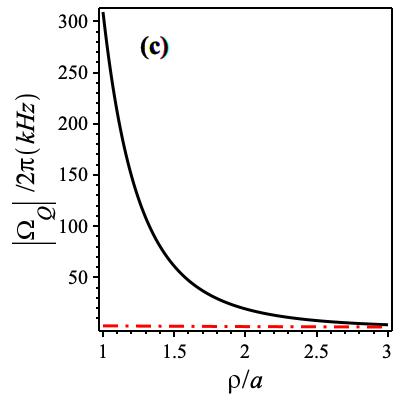}
\caption{(Color online) variations of the quadrupole Rabi frequency $|\Omega_Q|/2\pi (kHz)$ with $\rho$ for the quadrupole transition $\ket {L=0,m_l=0}\rightarrow \ket{L=2, m_l'}$ in Cs.  The plots are for different values of the propagation index $s=\pm 1$. (a-c) for the mode $C=$ \{HE$_{21}\}$. (a) For $m_l'=0$. (b) for $m_l'=+1$ (Black solid line) and for $m_l'=-1$ (Red dash-dotted line). (c) for $m_l'=+2$ (Black solid line) and $m_l'=-2$ (Red dash-dotted line) . The dashed line in all panels concerns the case $s=-1$, while the solid line is for $s=+1$. In all cases $ p=+1$. The power is ${\cal P} = 2.5$($\mu$W ); for other parameters see the text.}\label{Fig6} 
\end{figure}

In order to explore the effect of the direction of the propagation of light in the nano-fiber on the quadrupole Rabi frequency, we present in  Figure \ref{Fig6} (a-c), the radial variation of the quadrupole Rabi frequency  $|\Omega_Q|/2\pi (kHz)$ for the Cs atom with the quadrupole transition $\ket {L=0,m_l=0}\rightarrow \ket{L=2, m_l'}$ in the mode $C=$ \{HE$_{2 1}\}$ for $m_l'=0,\pm 1. \pm 2$ and different directions of propagation $s=\pm 1$.  So the quadrupole transition $\ket {L=0,m_l=0}\rightarrow \ket{L=2, m_l'=\pm 2}$  does not depend on the direction of the propagation, as has been shown previously. 

Whenever the index  $s$ has the value  $s=-1$ we find that the magnitude of the quadrupole Rabi frequency is smaller than that for the corresponding case where $s=+1$ for the cases ($m_l'=0, \pm 1$). 

\begin{figure}[ht]
\centering
\includegraphics[width=0.3\linewidth,height=0.3\linewidth]{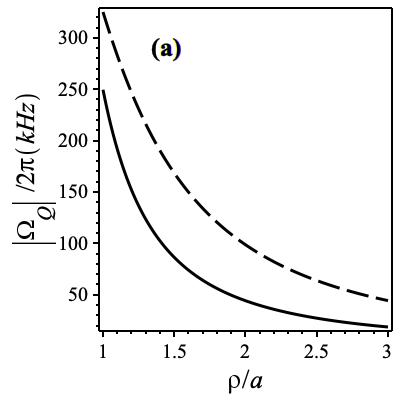}~\includegraphics[width=0.3\linewidth,height=0.3\linewidth]{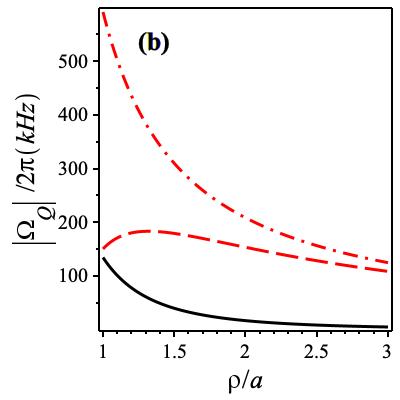} 
~\includegraphics[width=0.3\linewidth,height=0.3\linewidth]{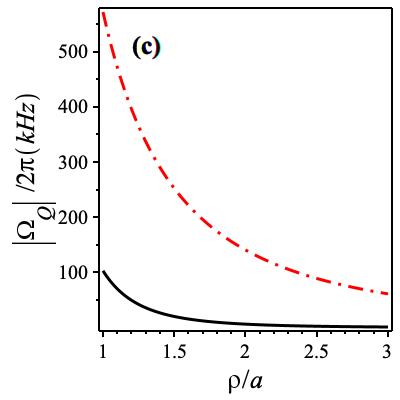}
\caption{(Color online) variations of the quadrupole Rabi frequency $|\Omega_Q|/2\pi (kHz)$ with $\rho$ for the Cs atom  quadrupole transition $\ket {L=0,m_l=0}\rightarrow \ket{L=2, m_l'}$.  The plots are for different values of the propagation $s=\pm 1$. (a-c) for the mode $C=$ \{HE$_{21}\}$. (a) For $m_l'=0$. (b) for $m_l'=+1$ (Black solid line) and for $m_l'=-1$ (Red dashed line). (c) for $m_l'=+2$ (black solid line) and $m_l'=-2$ (red dash-dotted line) . The dashed line in all panels is for the case $s=-1$, while the solid and dash-dotted lines are for $s=+1$. In all cases $ p=-1$. The power is ${\cal P} = 2.5$($\mu$W ); for other parameters see the text.}\label{Fig7} 
\end{figure}

It is also of interest to examine the effect of the polarization index $p$ of light in the nano fiber on the quadrupole Rabi frequency, so Fig. \ref{Fig7} shows the radial variations of the absolute value of the quadrupole Rabi frequency  $|\Omega_Q|/2\pi (kHz)$ for the Cs atom quadrupole transition $\ket {L=0,m_l=0}\rightarrow \ket{L=2, m_l'}$ in the mode $C=$ \{HE$_{2 1}\}$ for $m_l'=0,\pm 1,\pm 2$ and different value of the polarization parameter  $p=\pm 1$.  It is seen that the magnitude of the quadrupole Rabi frequency is affected by the polarization index $p$ and the maximum magnitude corresponds to the last panel for $m_l'=-2$, $s=\pm 1$, and $p=-1$.

The fact that we obtain different values for the Rabi frequency for different directions of propagation is a chirality effect, as described by the interesting work of Rauschenbeutel and coworkers \cite{PhysRevA.90.023805, petersen2014chiral, Lodahl_2017, Kien2017}.

The Rabi frequency in general shows a decrease with the radial position in all the curves shown here plotted with different parameters, except for the red dashed curve in 6(b).  This shows a  slight increase close to the fiber, leading to a shallow maximum, followed by a general decrease as for other curves.  This behavior can be explained with reference to Eq.(25).  It is seen that in the case where both s and p are negative, the general tendency to decrease controlled by the p term is initially dominated by the s-dependent term which then diminishes with the radial position.

Having evaluated the quadrupole Rabi frequency for the different cases allowed by the selection rules and the optical fiber parameters our final task is to evaluate the corresponding quadrupole transition rates.

\section{Transition amplitude and absorption rate}\label{sec5}

\begin{figure}[ht]
\centering
\includegraphics[width=0.3\linewidth,height=0.3\linewidth]{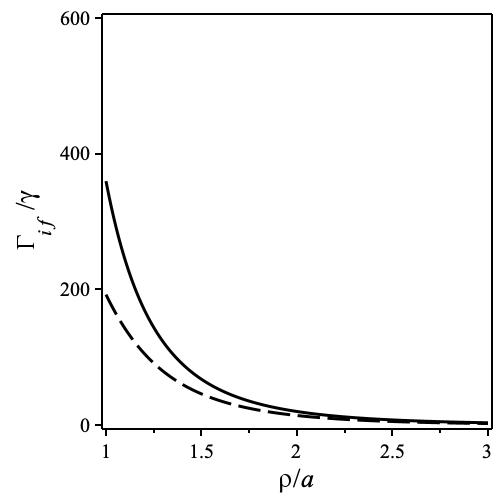}~\includegraphics[width=0.3\linewidth,height=0.3\linewidth]{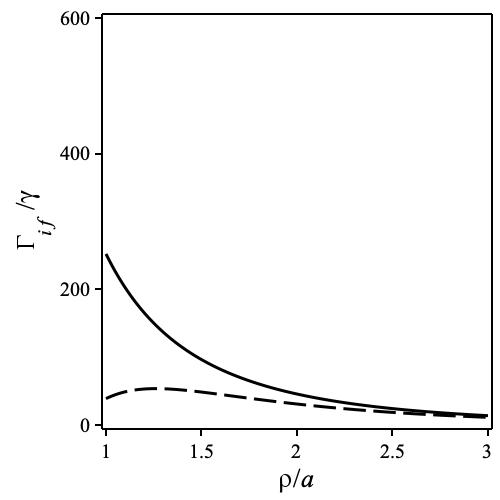} 
~\includegraphics[width=0.3\linewidth,height=0.3\linewidth]{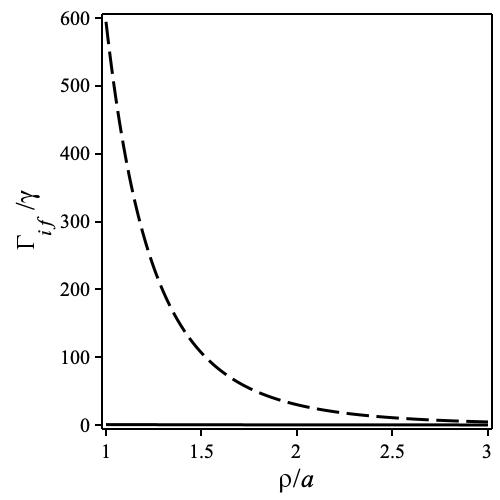}
\caption{Variations of the quadrupole absorption rate $\Gamma_{if}/\gamma$ for the Cs atom quadrupole transition $\ket {L=0,m_l=0}\rightarrow \ket{L=2, m_l'}$  with (a) $m_l'=0$, (b) $m_l'=- 1$, and (c) $m_l'=- 2 $ for the mode $C=$ \{HE$_{21}\}$. Solid lines $(s=+1, p=+1)$, dashed lines $(s=-1, p=-1)$.The power is ${\cal P} = 2.5$($\mu$W ); for other parameters see the text. All panels indicate a huge (more than two orders of magnitude) increase in the absorption rate for atomic radial position $\rho=a$.  This is attributable to the enhanced density of states just outside the fiber surface}\label{Fig8a} 
\end{figure}

The transition matrix element \cite{Bougouffa, ScholzMarggraf}, comprising only the quadrupole interaction, is $\mathrm{M}^{\{\alpha\}}_{if}=\bra{f}\hat{H}_{Q}\ket{i}$,
where $\ket{i}$ and $\ket{f}$ are, respectively, the initial and final states of the overall quantum system (atom plus fiber mode). We assume that the system has as an initial state $\ket{i}$ with the atom in its ground state and there is one optical fiber photon.  The final state $\ket{f}$  consists of the excited state of the atom and there is no field mode. Thus $\ket{i}=\ket{g\{1\}_{\{\alpha\}}}$ and $\ket{f}=\ket{e\{0\}}$.  We make use of the relations 
$\bra{\{0\}}\hat{a}^+_{\{\alpha'\}}\ket{\{1\}_{\{\alpha\}}}=0$ and $
\bra{\{0\}}\hat{a}_{\{\alpha'\}}\ket{\{1\}_{\{\alpha\}}}=\delta_{\{\alpha'\}\{\alpha\}}$,
where $a_{\{\alpha\}}$ and $a_{\{\alpha\}}^{\dagger}$ are the annihilation and creation operators of the fiber mode $\alpha$. We obtain
\begin{equation}\label{8}
\mathrm{M}^{\{\alpha\}}_{if}=\hbar \Omega_{Q}^{\{\alpha\}}(\mathbf{R})e^{i\theta^{\{\alpha\}}(\mathbf{R})}
\end{equation}
where  $ \Omega_{Q}^{\{\alpha\}}(\mathbf{R})$ is the quadrupole Rabi frequency. 
The final state of the system in the absorption process comprises a continuous band of energy of width $\hbar\gamma$ where $\gamma $ is the de-excitation rate involved in the quadrupole transition.  In this case, the absorption rate is governed by the form of Fermi's golden rule \cite{Bougouffa, Barnett2002} with a density of states 

\begin{eqnarray}\label{9}
    \Gamma_{if}= 2\pi\big | \Omega_{Q}^{\{\alpha\}}(\mathbf{R})\big|^2\mathcal{\rho}_{\omega_a}(\omega),
\end{eqnarray}
where $\mathcal{\rho}_{\omega_a}(\omega) $ is the density of the final state which is well represented by a Lorentzian distribution of states of width (FWHM) matching the free space spontaneous quadrupole emission rate.  Thus
\begin{equation}\label{9ppp}
\mathcal{\rho}_{\omega_a}(\omega)= \frac{1}{\pi}\frac{\gamma/2}{(\omega-\omega_a)^2+(\gamma/2)^2},
\end{equation}
The Lorentzian distribution characterizing the density of states specifies a limit to the validity of using Fermi's Golden rule to calculate the absorption rate since such a rate is valid only if the frequency width of the upper state $\ket{e}$ is larger than the excitation rate; i.e., the spontaneous emission rate is larger than the Rabi frequency.  The Rabi frequency may exceed the spontaneous emission rate for high intensities, in which case the perturbative approach culminating in the Fermi Golden Rule is no longer valid and the strong coupling regime is applicable involving Rabi oscillations.
Substituting Eq. (\ref{9ppp}) in Eq. (\ref{9}) we find for  the quadrupole absorption rate
\begin{equation}\label{main}
\Gamma_{if}=  \frac{\gamma}{(\omega-\omega_a)^2+(\gamma/2)^2}\big |\Omega_{Q}^{\{\alpha\}}(\mathbf{R})\big|^2
\end{equation}
This general expression applies to the various cases involving the different fiber modes participating in transitions satisfying the selection rules as discussed above. For illustration only, we display in Fig. \ref{Fig8a}, the variations of the absorption rate $\Gamma_{if}/\gamma$ with the radial position of the atom $\rho/a$  outside the fiber $\rho>a$.  We have concentrated on the quadrupole transition $\ket {L=0,m_l=0}\rightarrow \ket{L=2, m_l'}$ for different directions of propagation $s=\pm 1$ and the mode $C=$ \{HE$_{2 1}\}$. It can be seen for this particular case the absorption rate decreases with increasing $\rho$ and it is independent of the phase circulation direction $p=\pm 1$. Its magnitude close to the fiber surface $\rho\approx a$ can be more than two orders of magnitude relative to $\gamma$. The corresponding orbital angular momentum transfer rate in the case of $\Delta L=2$ is simply $2\hbar \Gamma_{if}$.

\section{Conclusions}\label{sec6}
In conclusion, we have systematized the formalism leading to the evaluation of the quadrupole absorption rate in the case of two-level systems localized outside the surface of an optical nano-fiber with a step-index profile, leading to modes that decay with radial distance outside the fiber.  The evaluation of the rates required, as a first step, the determination of the quadrupole Rabi frequencies for the hybrid modes in the optical fiber.  The quadrupole selection rules determine the form of the different types of field distribution that enters the coupling of the fields to the quadrupole moments.  Once the quadrupole Rabi frequencies are determined the absorption rate follows using the Fermi Golden rule with a Lorentzian final density of states function for the upper atomic level.

The hybrid modes are characterized by several special features that impact the evaluation of the absorption rates.  Firstly these modes influence the form of the quadrupole tensor components that enter a specific interaction with the electric field gradients.  Also, the absorption process involves a transaction between the fiber modes and the two-level system in which a quantum with specific characteristics is absorbed. Other characteristics of the fiber modes that affect the interaction are whether the mode is propagating along the +z-direction, or -z direction and whether the helical rotation is clockwise or anti-clockwise.   Such aspects will need to be taken into account in the context of actual experimental measurements. The observation that the results depend on the direction of propagation is interesting as has been pointed out recently by Ladahl et al \cite{Lodahl_2017, le2017chirality}) who, appropriately, referred to such a feature as a chirality effect.

We have focused on the quadrupole transition in  Cs primarily because this quadrupole transition has been the subject of recent investigations involving the interaction of Cs in the fields of optical fibers.  We have chosen to consider the simplest type of optical fiber with a uniform core material of refractive index $n_1>1$ immersed in free space.  Various other forms of optical fiber can be considered in which the core material is enclosed in a thin metallic layer, or a doped semiconductor in which case the dielectric function may be frequency- and wave vector-dependent and may have a complicated profile.  The uniform core fiber system discussed here has led to results for the absorption rate, as shown in Fig. 7, which should be experimentally measurable. Note that we have expressed the absorption rate $\Gamma_{if}$ in terms of $\gamma$ whose value is known for Cs and we have considered it a good scaling parameter to use in this context as done in a recent experiment \cite{Ray2020a}. Note also that we have considered only one specific value of the input power ${\cal P}$ used to excite the fiber mode with which the atom interacts.  Similar evaluations can be carried out for other value of ${\cal P}$.  From Fig.7  we can see that the absorption rate can be at least two orders of magnitude larger than $\gamma$ and are amenable to experimental measurements.

\section*{Acknowledgements}

The authors are grateful to Professors Stephen Barnett, Sile Nic Chormaic, and Fam Le Kien for useful correspondence.

\appendix

\section{Evaluation of ${\cal N}$}

The overall constant field amplitude ${\cal N}$ is fixed in terms of the experimentally controlled applied cycle-averaged field power ${\cal P}$ which is formally defined as the surface integral of the Poynting vector over a cross-section of the fiber.  To evaluate this in the case of the fiber modes we need expressions for the magnetic field components of the fiber nodes which are as follows 

{\underline {Hybrid modes}}:

 \begin{itemize}
  \item In the core region $0 \leq \rho \leq a$ we have:
  \begin{eqnarray}\label{e2h}
\mathcal{H}_{\rho}&=& i\mathcal{N}\frac{\omega\epsilon_0n_1^2}{2\mu}[(1-\sigma _1)J_{\ell-1}(\mu \rho)+(1+\sigma_1)J_{\ell+1}(\mu \rho],\nonumber\\
\mathcal{H}_{\varphi}&=& \mathcal{N}\frac{\omega\epsilon_0n_1^2}{2\mu}[(1-\sigma_1)J_{\ell-1}(\mu \rho)-(1+\sigma_1)J_{\ell+1}(\mu \rho)],\nonumber\\
\mathcal{H}_{Z}&=&- \mathcal{N}\frac{\beta}{ \omega \mu_0}\xi J_{\ell}(\mu \rho), 
\end{eqnarray}
  \item In the cladding (vacuum) region $\rho>a$ we have
\begin{eqnarray}\label{e3h}
\mathcal{H}_{\rho}&=& i\mathcal{N}\frac{\omega\epsilon_0n_2^2}{2\nu}\frac{J_{\ell}(\mu a)}{K_{\ell}(\nu a)}[(1-\sigma_2)K_{\ell-1}(\nu \rho)-(1+\sigma_2)K_{\ell+1}(\nu \rho)],\nonumber\\
\mathcal{H}_{\varphi}&=&i\mathcal{N}\frac{\omega\epsilon_0n_2^2}{2\nu}\frac{J_{\ell}(\mu a)}{K_{\ell}(\nu a)}[(1-\sigma_2)K_{\ell-1}(\nu \rho)-(1+\sigma_2)K_{\ell+1}(\nu \rho)],\nonumber\\
\mathcal{H}_{Z}& = &- \mathcal{N}\frac{\beta}{ \omega \mu_0}\xi\frac{J_{\ell}(\mu a)}{K_{\ell}(\nu a)}K_{\ell}(\nu \rho),
\end{eqnarray}
where 
\begin{eqnarray}\label{e4h}
\sigma_1=\frac{\beta^2}{k^2n_1^2}\xi, \quad \sigma_2=\frac{\beta^2}{k^2n_2^2}\xi.
\end{eqnarray}

\end{itemize}

The time-averaged Poynting vector component along the z-axis per unit area is expressed as
\begin{equation}
\label{e9p}
S_Z=\frac{1}{2}\Big(\mathbf{E}\times\mathbf{H}^*\Big)\cdot \hat{u}_Z=\frac{1}{2}\Big(\mathcal{E}_\rho\mathcal{H}_\varphi^*- \mathcal{E}_\varphi\mathcal{H}_\rho^*\Big),
\end{equation}
where $\hat{u}_Z$ is a unit vector in the Z-direction. The power residing in a mode of the optical fiber is then given by 

\begin{equation}
\label{e10p}
\mathcal{P} =\int_0^{2\pi}\, \int_0^{\infty} S_Z\,\rho\, d\rho\, d\varphi =\frac{1}{2} \int_0^{2\pi} \int_0^{\infty} \big(\mathcal{E}_\rho\mathcal{H}_\varphi^*- \mathcal{E}_\varphi\mathcal{H}_\rho^*\big)\,\rho\, d\rho\, d\varphi.
\end{equation}

The analytical expressions of the power flow for the hybrid modes are quite intricate. Here we give only the equations of the power as the sum of contributions from the core and cladding regions, respectively,

\begin{eqnarray}
\label{e13p}
\mathcal{P}_{core}&=&\frac{\pi}{4}\omega\epsilon_0n_1^2\beta|\mathcal{N}|^2\frac{a^2}{\mu^2}\Bigg [ (1-\xi)(1-\sigma_1)\int_0^{a}\,J_{\ell-1}^2(\mu \rho)\,\rho\,d\rho \nonumber \\
&+& (1+\xi)(1+\sigma_1)\int_0^{a}\,J_{\ell+1}^2(\mu \rho)\,\rho\,d\rho\Bigg] \nonumber\\
&=&|\mathcal{N}|^2{\cal I}_{H1}
\end{eqnarray}
\begin{eqnarray}
\mathcal{P}_{clad}&=&\frac{\pi}{4}\omega\epsilon_0n_2^2\beta|\mathcal{N}|^2\frac{a^2J_\ell^2(\mu a)}{\nu^2K_\ell^2(\nu a)}\Bigg [ (1-\xi)(1-\sigma_2)\int_a^{\infty}\,K_{\ell-1}^2(\nu \rho)\,\rho\,d\rho\nonumber \\
&+& (1+\xi)(1+\sigma_2)\int_a^{\infty}\,K_{\ell+1}^2(\nu \rho)\,\rho\,d\rho\Bigg]\label{e14p}\nonumber\\
&=&|\mathcal{N}|^2{\cal I}_{H2}
\end{eqnarray}
where $\sigma_1$ and $\sigma_2$ are given in terms of $\xi$ by Eq.(\ref{e4h}) and ${\cal I}_{H1,H2}$ are the rest of the expressions in $\mathcal{P}_{core,clad}$.

The total power is the sum
\begin{equation}
 \mathcal{P}=  \mathcal{P}_{core}+ \mathcal{P}_{clad}=|\mathcal{N}|^2({\cal I}_{H1}+{\cal I}_{H2})
\end{equation}
Thus we have
\begin{equation}
|\mathcal{N}|^2=\frac{{\mathcal{P}}}{{\cal I}_{H1}+{\cal I}_{H2}}
\end{equation}
The undetermined constant $\mathcal{N}$ can be determined when the total power flow $\mathcal{P}$ in optical fiber is specified.

\section{Radial variations of $S_{\varphi}$ for $\rho\geq a$}
\begin{figure}[ht]
\centering
\includegraphics[width=0.32\linewidth,height=0.4\linewidth]{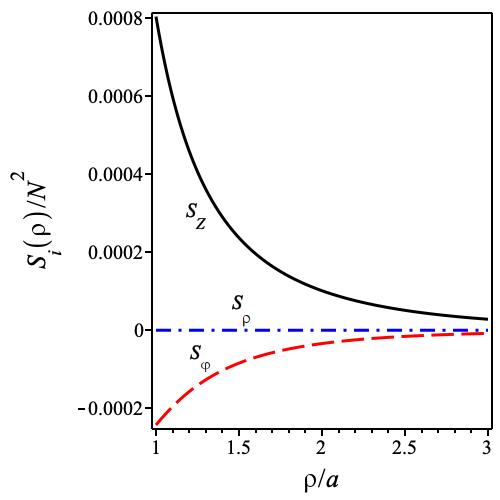}
\includegraphics[width=0.32\linewidth,height=0.4\linewidth]{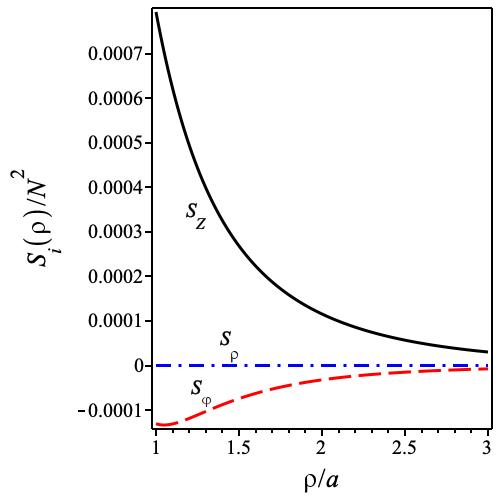}
\includegraphics[width=0.32\linewidth,height=0.4\linewidth]{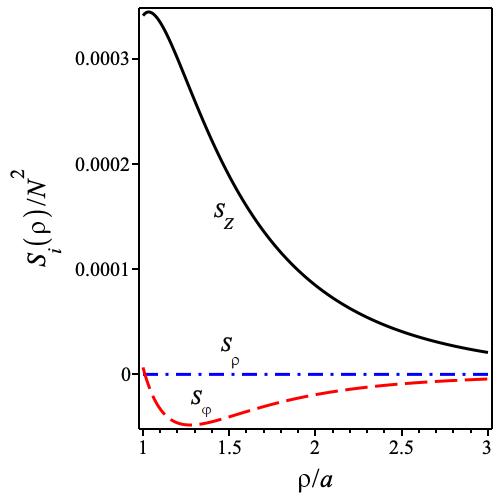}
\caption{(Color online) The radial variation of the real parts of the Poynting components for the mode $C=$ \{HE$_{21}\}$. The black solid line for $S_Z$.  The Red long-dashed line for $S_{\varphi}$. The blue dash-dotted line for $S_{\rho}$. 
Left panel: $a=290\ nm$, mid panel: $a=340\ nm$ and right panel: $a=400\ nm$. For Cs $\lambda=685\ nm$.
 For other parameters see the text.}\label{Fig8} 
\end{figure}

The time-averaged Poynting vector is 
\begin{equation}
\label{e9p }
\mathbf{S}=\frac{1}{2}\Re\Big(\mathbf{E}\times\mathbf{H}^*\Big).
\end{equation}
so its components in cylindrical coordinates are given as
\begin{eqnarray}
S_Z&=& \frac{1}{2}\Big(\mathcal{E}_\rho\mathcal{H}_\varphi^*- \mathcal{E}_\varphi\mathcal{H}_\rho^*\Big),\nonumber\\
S_{\varphi}&=& \frac{1}{2}\Big(\mathcal{E}_Z \mathcal{H}_\rho^*- \mathcal{E}_\rho\mathcal{H}_Z^*\Big),\nonumber\\
S_{\rho}&=& \frac{1}{2}\Big(\mathcal{E}_\varphi\mathcal{H}_Z^*- \mathcal{E}_Z\mathcal{H}_\varphi     ^*\Big),
\end{eqnarray}

where the factor $1/2$ that appears in the above expressions accounts for the time average of the Poynting vector. 
The electric and magnetic field components are given earlier, so the evaluations are straightforward, albeit somewhat cumbersome.  We, therefore, present the variations of the real parts of the Poynting vector components with radial variable $\rho$ specifically for the mode $C = \{HE_{21}\}$.  The results are shown in Fig(\ref{Fig8}).

It is clear that the radial component is practically zero and the azimuthal component is small compared to the z-component of the Poynting vector; i.e. $S_Z > 4 S_{\varphi}$ for different values of the fiber radius $a=290, 340, 400\ nm$. We may now conclude that the second term in Eq.(\ref{ell0}) is small, so we may assume ${\cal L}_z^{atom}={\cal L}_z^{fiber}$.

\bibliographystyle{apsrev4-2}
\bibliography{Ref2}

\end{document}